\documentstyle[prl,aps,psfig,epsf,twocolumn]{revtex}
\begin{document}
\draft
\twocolumn[\hsize\textwidth\columnwidth\hsize\csname@twocolumnfalse%
\endcsname

\preprint{}

\title{$O(4)$-Invariant Formulation
of the Nodal Liquid}
\author{Chetan Nayak}
\address{Physics Department, University of California, Los Angeles, CA 
  90095--1547}
\date{\today}
\maketitle

\begin{abstract}
We consider the $O(4)$ symmetric
point in the phase diagram of an electron
system in which there is a transition
between $d_{{x^2}-{y^2}}$ density-wave order
and $d_{{x^2}-{y^2}}$ superconductivity.
If the pseudospin $SU(2)\subset O(4)$
symmetry is disordered by quantum fluctuations,
the Nodal Liquid can result.
In this context, we
(1) construct a pseudospin $\sigma$-model;
(2) discuss its topological excitations;
(3) point out the
possibility of a {\it pseudospin-Peierls}
state and (4)
propose a phase diagram for the underdoped
cuprate superconductors.
\end{abstract}
\vspace{1 cm}
\vskip -0.5 truein

]
\narrowtext


{\it Introduction.}
Competing interactions and fluctuations
have led to a cornucopia of interesting
phenomena in the cuprate superconductors.
Unfortunately, these phenomena have not led to the
unambiguous determination of the phase diagram
of these materials, possibly because
some of the phases realized in these materials
are characterized by
particularly subtle forms of order.
This dilemma is rather
acute on the underdoped side of the
phase diagram, where it is still not clear
if the pseudogap can be ascribed to
a new phase of matter, a nearby critical point,
or a crossover. Since a better understanding
of proposed exotic phases and the transitions between them
may mitigate this difficulty, we study the
transition between the $d_{{x^2}-{y^2}}$
superconducting state of the cuprates
and a putative $d_{{x^2}-{y^2}}$
density-wave state (also known as the staggered flux
state \cite{staggered}; see \cite{Nayak00a}).
We ask if the pseudo-gap --
which appears to have $d_{{x^2}-{y^2}}$ symmetry
-- could be due to the proximity of the experimental
system to this transition.
The resulting phase diagram automatically
includes the Nodal Liquid state
\cite{Balents98,Balents99a,Balents99b},
a state with spin-charge separation.
We discuss the possible relevance of this theoretical cuprate
phase diagram to the experimental one.

{\it $O(4)$ Formulation of $d_{{x^2}-{y^2}}$
Ordered States at Half-Filling.}
In \cite{Nayak00a}, we adapted Yang's pseudospin
$SU(2)$ symmetry \cite{Yang89} to a critical point between
a $d_{{x^2}-{y^2}}$ density-wave state
and a $d_{{x^2}-{y^2}}$ superconductor.
The original pseudospin $SU(2)$
was germane to the transition between
a CDW and an $s$-wave superconductor;
Zhang's closely related $SO(5)$,
to the transition between an antiferromagnet and
a $d$-wave superconductor.

We first consider a transition at
half-filling between a singlet
commensurate $d_{{x^2}-{y^2}}$ density-wave
and a $d_{{x^2}-{y^2}}$ superconductor. We combine the order
parameters into
\begin{eqnarray}
\label{eqn:phi_def}
{\Phi_{\underline i}}(q)\,f(k) = \left(
\begin{array}{c}
\sqrt{2}\,{\rm Re}\left\{\left\langle
{\psi_\uparrow^\dagger}(k+\frac{q}{2})\,
{\psi_\downarrow^\dagger}(-k+\frac{q}{2}) \right\rangle\right\}\\
\sqrt{2}\,{\rm Im}\left\{\left\langle
{\psi_\uparrow^\dagger}(k+\frac{q}{2})\,
{\psi_\downarrow^\dagger}(-k+\frac{q}{2}) \right\rangle\right\}\\
i\left\langle {\psi^{\alpha\dagger}}(k+Q+\frac{q}{2})\,
{\psi_\alpha}(k-\frac{q}{2}) \right\rangle
\end{array}
\right)
\end{eqnarray}
where $f(k)=\cos{k_x}a-\cos{k_y}a$. 
Following Yang \cite{Yang89}, we introduce the
pseudospin $SU(2)$ generators ${O^3},{O^+}, {O^-}=({O^+})^\dagger$
\begin{eqnarray}
\label{eqn:pseduo-def}
{O^3} &=& {\int_{\rm R.B.Z.}}\frac{{d^2}k}{(2\pi)^2}\,\,
\biggl({\psi^{\alpha\dagger}}(k)\,
{\psi_\alpha}(k)\,\, +\,\,\,
k\rightarrow k+Q
\biggr)\cr
{O^+} &=& {\int_{\rm R.B.Z.}}\frac{{d^2}k}{(2\pi)^2}\,\,
i{\psi_\uparrow^\dagger}(k)\,
{\psi_\downarrow^\dagger}(-k+Q)
\end{eqnarray}
The order parameters form a triplet
under this $SU(2)$\footnote{We will use underlined lowercase
Roman letters such as
${\underline i}={\underline 1},{\underline 2},{\underline 3}$
to denote pseudospin triplet indices
and uppercase Roman letters to denote peudospin doublet
indices $A=1,2$. Lowercase Roman indices $a=1,2,3$
will be vector indices (i.e. real spin triplet
indices) and Greek letters $\alpha=1,2$ will
be used for real spin $SU(2)$ spinor indices.
Pauli matrices $\tau^{\underline i}$ will be used
for pseudospin, while $\sigma^a$ will be reserved for
spin.}. The integrals are over the reduced Brillouin zone.

There is a small but important difference
between our pseudospin $SU(2)$ and
Yang's \cite{Yang89}: the factors of
$i$ in ${O^\pm}$.
They are necessary since a commensurate
$d_{{x^2}-{y^2}}$ density-wave
breaks $T$, while a superconductor
does not; hence, our pseudospin $SU(2)$
does not commute with $T$. Pseudospin $SU(2)$,
spin $SU(2)$, and time-reversal combine to
form the symmetry group $O(4)$.

The electron fields transform as
a doublet under both $SU(2)$s. We will
group them into $\Psi_{A\alpha}$:
\begin{eqnarray}
\left(
\begin{array}{c}
{\Psi_{1\alpha}}\\ {\Psi_{2\alpha}}
\end{array}
\right)
= \left(
\begin{array}{c}
{\psi_\alpha}(k)\\
i{\epsilon_{\alpha\beta}}{\psi^{\beta\dagger}}(-k+Q)
\end{array}
\right)
\end{eqnarray}

Near the transition between a
$d_{{x^2}-{y^2}}$ density-wave
and a $d_{{x^2}-{y^2}}$ superconductor, we
can focus on the low-energy degrees of freedom:
the order parameters and the nodal
fermionic excitations.
We can write down an $O(4)$-invariant action
for this transition:
\begin{eqnarray}
\label{eqn:o(4)_Lag}
{S_{\rm eff}} &=&
 \int d\tau\,\frac{{d^2}k}{(2\pi)^2}\,
{\Psi^{A\alpha^\dagger}} \left({\partial_\tau}
 -  \epsilon(k)\right) {\Psi_{A\alpha}}\,+\cr
& &{\hskip -1 cm}i\,g \int d\tau\,
\frac{{d^2}k}{(2\pi)^2}\,\frac{{d^2}q}{(2\pi)^2}
\,\,{\Phi_{\underline i}}(q)\,f(k)\,\times\cr
& &\Bigl[
{\epsilon^{\alpha\beta}}{\Psi_{C\alpha}}\left(k+\frac{q}{2}\right)
{\epsilon^{CA}}
{\tau^{\,{\underline i}\,B}_A} {\Psi_{B\beta}}
\left(-k+\frac{q}{2}\right)\,+\cr
& &{\hskip 0.2 cm}{\epsilon_{\alpha\beta}}
{\Psi^{A\alpha\dagger}}\left(k+\frac{q}{2}\right)
{\tau^{\,{\underline i}\,B}_A} {\epsilon^{BC}}
{\Psi^{B\beta\dagger}}\left(-k+\frac{q}{2}\right)\Bigl] \cr
& &{\hskip -0.6 cm}+\,\int d\tau {d^2}x\,\left(
{\left({\partial_\mu}{\Phi_{\underline i}}\right)^2}+
\frac{1}{2}\,r\,{\Phi_{\underline i}}{\Phi_{\underline i}}
+ \frac{1}{4!}\,u\,
{\left({\Phi_{\underline i}}{\Phi_{\underline i}}\right)^2}
\right)
\end{eqnarray}
`Microscopic' models with this symmetry were
constructed in \cite{Nayak00a}.
In this $O(4)$-symmetric action, we have, by a rescaling, set the
${\Phi_{\underline i}}$ velocities, $v_{\underline i}$,
and stiffnesses, ${\rho_{\underline i}}$, to $1$.
This cannot be done in the asymmetric case,
${\rho_{\underline 1}}={\rho_{\underline 2}}\equiv{\rho_s}
\neq {\rho_{DW}}\equiv{\rho_{\underline 3}}$.
In general, symmetry-breaking terms will be present,
but they can scale to zero at a critical point,
thereby dynamically restoring the symmetry,
as we discuss later. Hence, we focus on the symmetric case.

When ${\Phi_{\underline i}}$ is
ordered, the fermionic spectrum is
$E(k) = \sqrt{{\epsilon^2}(k) +
{g^2}{\Phi_{\underline i}}{\Phi_{\underline i}}{f^2}(k)}$.
In the following, we ignore the fermionic excitations
which are not associated with the nodes
of the $d_{{x^2}-{y^2}}$ order parameter.
We linearize $\epsilon(k)$ about the Fermi surface
and $f(k)$ about the nodes.
If we rotate our axes so that the $k_x$ axis is perpendicular
to the Fermi surface at one antipodal pair of nodes,
then we can write $\epsilon(k)\approx{v_F}{k_x}$
and $g\left|{\Phi_{\underline i}}\right|f(k)\approx
{v_\Delta}{k_y}$.
As in \cite{Balents98}, we will have to introduce an additional
index $a=1,2$ for the two sets of antipodal nodes
which differ by the replacement
${k_x}\leftrightarrow{k_y}$. In order to avoid unnecessary
clutter, this index will be suppressed.

It is convenient to adopt a non-linear $\sigma$-model
approach and assume that the magnitude
of ${\Phi_{\underline i}}$ is fixed, ${\Phi_{\underline i}^2}={a^2}$.
Following \cite{Shraiman89,Sachdev94}, we
employ a $CP^1$ representation
of the non-linear $\sigma$ model:
\begin{equation}
{\Phi^{\underline i}} = {z^{A\dagger}}{\tau^{\,{\underline i}
\,\,B}_{\,A}}{z_B}
\end{equation}
with ${\left|{z_1}\right|^2} + {\left|{z_2}\right|^2} = a^2$
and rotate the pseudospins of the fermions to the
local direction of the order parameter:
\begin{equation}
\label{eqn:chi_def}
{\Psi_A} = {U_A^B}{\chi_B}
\end{equation}
where
\begin{eqnarray}
U = \frac{1}{a}\,\left(
\begin{array}{cc}
{z_1} & -{z_2^*} \\ 
{z_2} & {z_1^*}
\end{array}
\right)
\end{eqnarray}
The latter change of variables
is a direct $SU(2)$ analogue of the original
$U(1)$ Nodal Liquid construction
\cite{Balents98}.
As in that case, it is
is double-valued, so we must introduce
a Chern-Simons term as in
\cite{Balents99a} which couples
the $\chi$'s to the topological current
${j_\mu} =  {\epsilon_{\mu\nu\lambda}}
{\epsilon_{{\underline i}{\underline j}{\underline k}}}
{\Phi^{\underline i}}
{\partial^\nu}{\Phi^{\underline j}}
{\partial^\lambda}{\Phi^{\underline k}}$.
This term is only important at the phase
transitions since the topological current
vanishes in the ordered phases
since the pseudospins are aligned
and in the disordered phases
since it is odd
under the $Z_2$ symmetry
${\Phi^{\underline 3}}\rightarrow -{\Phi^{\underline 3}}$.
We suppress this term below.

In terms of $z_A$, $\chi_A$, the effective
action takes the form:
\begin{eqnarray}
\label{eqn:o(4)_NL_action}
{S_{\rm eff}} &=&  \int d\tau\,{d^2}x\,
{\chi^{A\alpha\dagger}} \left({\partial_\tau}
+ {\alpha_\tau}{\tau^{\underline 3}} 
 - {v_F}\,i {\partial_x}
-{v_F}{\alpha_x}{\tau^{\underline 3}} 
 \right) {\chi^{}_{A\alpha}}\cr
& &{\hskip -0.5 cm}+ \,i\, \int d\tau\,\frac{{d^2}k}{(2\pi)^2}\,
\Bigl[{\epsilon^{\alpha\beta}}{\chi^{}_{C\alpha}}
{\epsilon^{CA}}
{\tau^{{\underline 3}\,B}_A} \,{v_\Delta}i{\partial_y}
{\chi^{}_{B\beta}}\,+\cr
& &{\hskip 2.4 cm}{\epsilon_{\alpha\beta}}
{\chi^{A\alpha\dagger}}
{\tau^{{\underline 3}\,B}_A} {\epsilon^{BC}}
{v_\Delta}i{\partial_y}{\chi^{B\beta\dagger}}\Bigl] \cr
& &{\hskip - 1 cm} +\,\int d\tau\,\frac{{d^2}k}{(2\pi)^2}\,\,
{\chi^{\alpha\dagger}} \Biggl({U^\dagger}
\left({\partial_\tau}-{A_\tau}
{\tau^{\underline 3}} \right)U -
{\alpha_\tau}{\tau^{\underline 3}} 
\cr & &
{\hskip 1.0 cm}
-\, {v_F}\,{U^\dagger}\left(i {\partial_x}-
{A_x}{\tau^{\underline 3}} \right)U
+ {v_F}{\alpha_x}{\tau^{\underline 3}} 
\Biggr)
{\chi^{}_{\alpha}}\cr
& &{\hskip - 1 cm}+\,\int d\tau {d^2}x\,\left(
{\left|\left(i{\partial_\mu} - {\alpha_\mu}
- {A_\mu}{\tau^{\underline 3}}
\right)z\right|^2} +
\lambda\left({z^\dagger}z - a\right)\right)
\end{eqnarray}
The $U(1)$ gauge field $\alpha_\mu$ is a Lagrange multiplier
which removes the redundant phase variable
in the parametrization of $CP^1$ by $z_A$.
A coupling between $\alpha_\mu$ and $\chi_A$
has been added to the first term
and subtracted from the ${U^\dagger}\partial U$
terms so as to make the latter invariant
under the gauge transformation
${z_A}\rightarrow{e^{i\theta}}{z_A}$.
$\lambda$ is a Lagrange multiplier
which fixes ${\Phi_{\underline i}^2}={a^2}$.
We have introduced the external
electromagnetic field, $A_\mu$,
in order to keep track of
the charge quantum numbers of the fields.
When $a$ is large, ${\Phi^{\underline i}}=
{z^{A\dagger}}{\tau^{\,{\underline i}
\,\,B}_{\,A}}{z_B}$ condenses and
the system is in one of the
$d_{{x^2}-{y^2}}$ ordered states. When $a$ is small,
${\Phi^{\underline i}}$ is disordered.
There is a critical point at $a={a_c}$.

{\it The Nodal Liquid Revisited.}
In the ordered phases, $U$ is a constant, so the
${U^\dagger}\partial U$ terms in (\ref{eqn:o(4)_NL_action})
can be dropped; the nodal quasiparticles
are coupled to the external
electromagnetic field.
Note that the $d_{{x^2}-{y^2}}$ density
wave is an ordered state in this formalism,
unlike in \cite{Balents98,Balents99a,Balents99b},
where it is a disordered state.
In the disordered phases,
the $z_A$ sector of the theory develops
a gap. Hence, the fourth and
fifth lines of (\ref{eqn:o(4)_NL_action})
can be dropped at low
energies. To analyze these phases further,
we introduce a dual representation for
$z_A$, following \cite{Balents99b,Nayak99}.
The effective action now takes the form:
\begin{eqnarray}
\label{eqn:o(4)_NL_dualaction}
{S_{\rm eff}} &=&  {S_F}[{\chi_A},{\alpha_\mu}] +
\,{\sum_A}{S_{GL}}\left[{\Phi^A},
\frac{1}{2}({a^+_\mu}\pm{a^-_\mu})\right]\cr
& & + \int d\tau\,{d^2}x\,\left(
{\alpha_\mu}{\epsilon_{\mu\nu\lambda}}{\partial_\nu}
{a^+_\lambda}
+ {A_\mu}{\epsilon_{\mu\nu\lambda}}{\partial_\nu}
{a^-_\lambda}\right)
\end{eqnarray}
where ${S_F}[{\chi_A},{\alpha_\mu}]$ is the first three
lines of (\ref{eqn:o(4)_NL_action}),
$\Phi^A$ annihilates a vortex in $z_A$, and
\begin{equation}
{{\cal L}_{GL}}({\Phi},{a_\mu}) = 
{1 \over 2} {|(i{\partial_\mu} - {a_\mu}){\Phi} |^2}
+ V({\Phi}) + {1 \over 2} {({f_{\mu \nu}})^2}
\end{equation}
and ${J^{\pm}_\mu} =
{\epsilon_{\mu\nu\lambda}}{\partial_\nu} {a^{\pm}_\lambda}$
are the $z_A$ number and pseudospin ${\underline 3}$
currents. When the $Z_2$ symmetry
${\Phi^A}\rightarrow - {\Phi^A}$ is unbroken,
we can rewrite the effective action
in terms of the fields
${\Phi^+}={\Phi^1}{\Phi^2}$,
${\Phi^-}={\Phi^1}{\Phi^{2\,\dagger}}$.
We now have:
\begin{eqnarray}
{S_{\rm eff}} &=&  {S_F}[{\chi_A},{\alpha_\mu}] +
{S_{GL}}[{\Phi^+},{a^+_\mu}] +
{S_{GL}}[{\Phi^-},{a^-_\mu}]\cr
& & + \,\int d\tau\,{d^2}x\,\left(
{\alpha_\mu}{\epsilon_{\mu\nu\lambda}}{\partial_\nu}
{a^+_\lambda}
+ {A_\mu}{\epsilon_{\mu\nu\lambda}}{\partial_\nu}
{a^-_\lambda}\right)
\end{eqnarray}
Integrating out $\alpha_\mu$, we can solve
the resulting constraint to express
${a^+_\mu}$ in terms of $\chi_A$:
${J^+_0}={\chi^\dagger}{\tau^{\underline 3}}\chi$,
${J^+_x}={v_F}{\chi^\dagger}{\tau^{\underline 3}}\chi$.

Now suppose that the system becomes disordered
as a result of the condensation of $\Phi^-$. By
the Anderson-Higgs mechanism, $a^-_\mu$
aquires a gap. Integrating out $a^-_\mu$,
we find no coupling of $A_\mu$ to the remaining
degrees of freedom: $\chi_A$ is a neutral
spin-$1/2$ fermion.
The change of variables (\ref{eqn:chi_def})
has effectively `bleached' the fermions by using the order parameter
to screen their pseudospin (including their charge).
This state is none other than the Nodal Liquid.

{\it Pseudospin-Peierls Order.}
If the system is, instead, disordered
by the condensation of $\Phi^{1,2}$,
then ${J^\pm_\mu}$ must vanish at low energies.
The only allowed excitations at low energies
are those combinations of $\chi_A$s which
are invariant under ${\tau^{{\underline 3}\,B}_A}$
rotations, i.e. neutral excitations. At finite energy, there are also
solitonic excitations which carry one quantum
of $({a^+_\mu}\pm{a^-_\mu})/2$ flux, i.e. charge $e$
and spin-$1/2$. According to the analogy between
the pseudospin $SU(2)$ physics of our system
and the spin $SU(2)$ physics of a
quantum antiferromagnet,
we might, in this disordered phase,
expect the pseudospin analog
of spin-Peierls order,
{\it pseudospin Peierls} order,
\begin{eqnarray}
\label{eqn:SP_ord}
\left\langle  \vec{\Phi}(k+K)\,\cdot\,
\vec{\Phi}(k) - 
\vec{\Phi}\times{\partial_\tau}\vec{\Phi}(k+K)
\cdot \vec{\Phi}\times{\partial_\tau}\vec{\Phi}(k)
\right\rangle\cr
= \sin{k_x}a
\end{eqnarray}
with $K=(\pi/a,0)$ or $(0,\pi/a)$,
as a result of Berry phases \cite{Read89b}
which we have neglected in (\ref{eqn:o(4)_NL_action}).

{\it Phase Transitions at Half-Filling.}
The transition at half-filling
between the $d_{{x^2}-{y^2}}$ density-wave
and the $d_{{x^2}-{y^2}}$ superconductor
is driven by a pseudospin-$2$ symmetry-breaking field,
\begin{eqnarray}
{S_u} = u\,\int d\tau\,{d^2}x\,
\left( {\Phi_3^2} - {\Phi_1^2} - {\Phi_2^2}\right)
\end{eqnarray}
For $u<0$, the ${\underline 3}$-axis is an easy axis
and the $d_{{x^2}-{y^2}}$ density-wave state is
favored; for $u>0$, the ${\underline 1}-{\underline 2}$-plane
is an easy plane and the $d_{{x^2}-{y^2}}$
superconducting state is favored. At $u=0$,
a first-order {\it pseudospin-flop} transition occurs,
provided $a>{a_c}$. At the bicritical point
$a={a_c}$, $u=0$, quantum fluctuations destroy
order at the $O(4)$-symmetric point.
This bicritical point and the quantum critical region
\cite{Chakravarty89,Sachdev99} are described by the physics
of the critical fluctuations coupled
to nodal fermionic excitations.
For $a<{a_c}$, $u=0$ the system
lies along the $O(4)$-symmetric line
in the Nodal Liquid phase. A small increase
or decrease of $u$ will not cause order,
and the system will still be in the nodal
liquid phase, albeit with lower
symmetry, $U(1)\times {Z_2}$. Further
increase or decrease of $u$ will lead to
second-order phase transitions at ${u^\pm_{cr}}(a)$
into the $d_{{x^2}-{y^2}}$ superconducting and $d_{{x^2}-{y^2}}$
density-wave phases respectively.

At the second-order transition from the Nodal Liquid to
the $d_{{x^2}-{y^2}}$ density-wave,
the $Z_2$ symmetry of translation by
one lattice site is broken. At the
second-order $XY$ transition from the Nodal Liquid to the
$d_{{x^2}-{y^2}}$ superconductor,
electromagnetic $U(1)$ is broken.
At the first-order pseudospin flop transition
between the $d_{{x^2}-{y^2}}$ superconductor
and the $d_{{x^2}-{y^2}}$ density-wave,
$U(1)$ is restored and $Z_2$
is simultaneously broken. 
In the formulation discussed here,
spin-charge confinement -- which,
in the language of \cite{Balents99a,Balents99b,Senthil99}
(see also \cite{Read91}) is due to
the absence of vortex
pairing -- occurs simultaneously
with translational symmetry breaking.

{\it Topological Excitations.}
We can give a narrative for the destruction of
superconductivity in
the language of vortex condensation.
In the superconducting
phase, the pseudospin
${\Phi^{\underline i}}$ lies in the
${\underline 1}-{\underline 2}$
plane. In the core of a vortex -- a {\it meron} in
the $\sigma$-model -- ${\Phi^{\underline i}}$
must point out of the ${\underline 1}-{\underline 2}$
plane. This can be done by pointing along
the $\pm {\underline 3}$ axis. When $+{\underline 3}$ merons
dominate (in the presence of
an infinitesimal $Z_2$ symmetry-breaking field),
the superconductor undergoes a transition to
the $d_{{x^2}-{y^2}}$ density-wave state.
When there are equal numbers of $\pm {\underline 3}$ merons,
the superconductor instead undergoes a transition to the disordered
state. This condition on the densities of
$\pm {\underline 3}$ merons is reminiscent of and cognate to
the vortex-pairing scenario of
\cite{Balents99a,Balents99b}, but is weaker since it
allows for the two possibilities discussed earlier.
The transition from the $d_{{x^2}-{y^2}}$ density-wave
state to the disordered state can be understood in terms of
{\it skyrmion} condensation.

\begin{figure}[htb]
\centerline{\psfig{figure=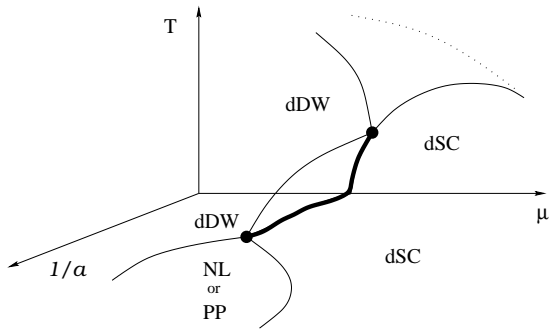,height=1.7in}}
\vskip 0.5cm
\caption{The proposed phase diagram for the cuprates.
The thick lines are first-order phase transitions;
the thin one, second-order. The large dots are bicritical
points, as is the entire thin line connecting them.}
\label{fig:T-mu-a_phase}
\end{figure}

{\it Discussion.}
Transitions of the type which we have
discussed above do not in the cuprates occur
at half-filling but -- if at all -- near $x_c$,
the doping at which superconductivity
first appears. We assume $u<0$ to suppress
superconductivity at half-filling.
In order to move away from
half-filling, we vary the chemical potential,
which can be done by adding the $O(4)$-breaking term:
\begin{eqnarray}
\label{eqn:chem_pot}
{S_{\mu}} = \mu\, {O^3}
= \mu\int d\tau\,{d^2}x\,\left(
{\epsilon_{{\underline 3}{\underline i}{\underline j}}}
{\Phi_{\underline i}}{\partial_\tau} {\Phi_{\underline j}}
+ {\Psi^\dagger} {\tau^{\underline 3}} \Psi \right)
\end{eqnarray}
By increasing $\mu$, we can drive the system
through a first-order pseudospin-flop transition
into the superconducting state.
As $a$ is decreased, a bicritical point will
again be reached. The coupling between $z^{}_A$
and $\chi^{}_A$ only enters at two-loops; at one-loop,
we can appeal to known results for the
pure non-linear $\sigma$-model,
which indicate that the $O(4)$ symmetry
is dynamically restored at the bicritical point
\cite{Pelcovits76}. As a result,
the $O(4)$-symmetric critical theory \cite{Chubukov94} discussed above
will apply in the low-frequency, long-wavelength
limit. A possible phase diagram for the
cuprates, based on this scenario,
is depicted in figure \ref{fig:T-mu-a_phase}.
An alternative, not depicted in figure \ref{fig:T-mu-a_phase},
can occur if ${\rho_s}<\rho_{DW}$. In this case,
there can be a phase with both $d_{{x^2}-{y^2}}$ superconducting and
$d_{{x^2}-{y^2}}$ density-wave order, and a
tetracritical point, $T={T_{bc}}$, $\mu={\mu_{bc}}$,
at which both orders become critical. For
$\mu<{\mu_{bc}}$, there will be a regime,
${T_c^{sc}}<T<{T_c^{dw}}$, above the superconducting
transition temperature, which has density-wave order.

The dotted line in figure \ref{fig:T-mu-a_phase}
is the pseudogap scale,
which we interpret as the
scale below which ${\Phi^{\underline i}}$
has fixed magnitude and the
non-linear $\sigma$ model description
is available.
Let us consider the physics below this scale.
As $\mu$ is increased, Fermi pockets open
at the nodes of the $d_{{x^2}-{y^2}}$ density-wave state.
Eventually, the system
undergoes a transition from the
$d_{{x^2}-{y^2}}$ density-wave to the
$d_{{x^2}-{y^2}}$ superconductor.
The nature of this transition depends on
the value of $a$ which, ostensibly, varies among
the materials in the cuprate family.
It may, perhaps, be controlled by chemical substitution
or applied pressure. For $a$ large,
the transition will be first-order
as depicted by the thick line.
For $a$ small, it occurs via two
second-order phase transitions;
the Nodal Liquid is sandwiched between
these two transitions. Neither the $d_{{x^2}-{y^2}}$ superconductor
nor the Nodal Liquid has Fermi pockets, the
latter because the second term in (\ref{eqn:chem_pot}) can
be dropped in the disordered phase. Appealing to the phase
diagram of the spin-flop transition in
magnetically-ordered systems, we extend the first-order
phase transition to finite-temperature, where
it meets the second-order $d_{{x^2}-{y^2}}$ density-wave
and superconducting ordering transitions.

Whither the antiferromagnet?
As Hsu \cite{Hsu90} and Gros \cite{Gros89}
pointed out, the $d_{{x^2}-{y^2}}$ density-wave
state has good short-ranged antiferromagnetic
correlations, reflected in its excellent numerical variational energy.
Hence, we will assume that the only additional physics
needed to describe the antiferromagnetic state at half-filling
is a moderate triplet quasiparticle-quasihole condensate \cite{Hsu90}.
This will not affect our
description of the critical regime.
Our assumption appears to be supported
by photemission experiments on the antiferromagnetic
insulator, ${Ca_2}Cu{O_2}{Cl_2}$ \cite{Ronning98}.
Similar ideas may apply to the
Nodal Liquid state, making it an equally good
platform for the antiferromagnetic state at half-filling.

Our non-linear $\sigma$-model analysis
mirrors that of \cite{Zhang97}, but is
on firmer footing because the
$d_{{x^2}-{y^2}}$ density-wave -- unlike
the antiferromagnet --
has a nodal fermionic spectrum
similar to that of the $d_{{x^2}-{y^2}}$ superconductor
into which the pseudospin symmetry rotates it.
Fluctuations between the
$d_{{x^2}-{y^2}}$ density-wave
and superconducting states
are also a key feature of the
$SU(2)$ mean-field-theory
of the $t-J$ model\cite{Wen96}. In fact,
a parallel approach
to the Nodal Liquid state was taken
in this framework in \cite{Wu99}.
However, the $SU(2)$ is local
in that approach, which leads
to complications arising from
the concomitant gauge field.
One virtue of the non-linear $\sigma$-model
approach is that we
can use the physics of quantum
antiferromagnets as a guide.
In this way, we identified
{\it pseudospin-Peierls} order
as a possible alternative to the
Nodal Liquid phase.
Another striking upshot
of our analysis is the bicritical
point at which the $d_{{x^2}-{y^2}}$ density-wave,
$d_{{x^2}-{y^2}}$ superconducting, and
Nodal Liquid phases touch. It is possible that
it is responsible
for recent experimental hints of quantum critical behavior
in the cuprates \cite{Aeppli97,Ando97}.

I would like to thank S. Chakravarty for
discussions, and S. Sachdev and
T. Senthil for pointing out an error in an earlier
version of this paper.


{

\end{document}